\documentclass[a4paper,10pt,twoside]{cpc-hepnp}
\usepackage{CJK,upgreek,fancyhdr}
\usepackage{multicol}
\usepackage{graphicx}
\usepackage{amssymb,bm} 
\usepackage[tbtags]{amsmath}
\usepackage{xcolor}              %

\def\bea{\begin{eqnarray}} \def\eea{\end{eqnarray}}
\def\be{\begin{equation}} \def\ee{\end{equation}}
\def\bal{\begin{align}} \def\eal{\end{align}}
\def\bse{\begin{subequations}} \def\ese{\end{subequations}}

\def\al{\alpha}
\def\eps{\varepsilon}
\def\ms{M_\odot}
\def\bds{B_\text{DS}}

\def\al{\alpha}
\def\rob{\rho_B}
\def\mmax{M_\text{max}}
\def\fm3{$\,\text{fm}^{-3}$}
\def\mfm{$\,\text{MeV}\,\text{fm}^{-3}$}

\begin{document}

\begin{CJK*}{GB}{gbsn}
\fancyhead[c]{\small Chinese Physics C~~~Vol. xx, No. x (201x) xxxxxx}
\fancyfoot[C]{\small 010201-\thepage}

\footnotetext[0]{Received 7 February 2017}
\title{
Two-flavor hybrid stars with the Dyson-Schwinger quark model}

\author{%
      J.-B. Wei (κ½ð±ê)$^{1}$
\quad H. Chen (³Â»¶)$^{1;1)} $\email{huanchen@cug.edu.cn}
\quad H.-J Schulze$^{2}$
}

\maketitle

\address{%
$^1$ School of Mathematics and Physics, China University of Geosciences,
Lumo Road 388, 430074 Wuhan, China\\
$^2$  INFN Sezione di Catania, Dipartimento di Fisica,
Universit\'a di Catania, Via Santa Sofia 64, 95123 Catania, Italy\\
}

\begin{abstract}%
We study the properties of two-flavor quark matter in the Dyson-Schwinger model
and investigate the possible consequences for hybrid neutron stars,
with particular regard to the two-solar-mass limit.
We find that with some extreme values of the model parameters,
the mass fraction of two-flavor quark matter in heavy neutron stars
can be as high as 30 percent
and the possible energy release during the conversion
from nucleonic neutron stars to hybrid stars can reach $10^{52}\;$erg.
\end{abstract}



\end{CJK*}
\begin{multicols}{2}

\section{Introduction}

The interior structure of massive neutron stars (NS)
is one of the main issues in the physics of compact stars \cite{Glen00}.
Recent observations confirm the existence of two NS of about
two solar masses \cite{Demorest10,Antoniadis13,Fonseca16}.
Based on a microscopic nucleonic equation of state (EOS),
one expects that in such heavy NS the central particle density
may reach values larger than $1/\text{fm}^3$,
where in fact quark degrees of freedom are expected to appear
at a macroscopic level.
There have been numerous studies
(see, e.g., \cite{Buballa14,Weber14,Orsaria14,Drago14,Castillo16,Alford16})
to demonstrate the possibility of quark matter (QM) in massive NSs.

However, there are still many open questions due to lack
of our knowledge on dense matter.
One issue is
the competition of QM with other degrees of freedom, such as hyperons;
another issue is
the realization and composition of QM in compact stars.
One possibility is based on the hypothesis of absolutely stable
strange quark matter (SQM),
which corresponds to the so-called strange quark star (SQS);
another possibility is that QM only appears in the core of a hybrid NS,
with a phase transition from hadronic matter (HM) to QM with three flavors
($u,d,s$), simply called 3QM in the following.
This is because the current mass of strange quarks is only about 100 MeV
and usually one expects a chiral symmetry restoration
at such a high density in compact stars.

However, it is also possible that strange quarks are suppressed
by their large effective mass
or by flavor conservation on a short time scale \cite{Dai95,Drago07}.
There are also works \cite{Klahn07} showing that the onset of the strange quark
flavor in a CFL phase marks an instability of the stars,
so that no stable 3QM-core stars might be found in nature.
It is therefore also interesting to investigate the possibility
that QM with only two flavors $u,d$ (2QM) appears in the core of NS.
It can be regarded as a limiting case of the 3QM hybrid stars
or a transitory step of the conversion from NS to SQS
\cite{Dai95,Drago07,Furusawa16a}.
We therefore present this work as a complement to our previous study
of 3QM hybrid stars \cite{Chen11,Chen15a,Chen16}.

The mass of a NS can be calculated by solving the Tolman-Oppenheimer-Volkoff
(TOV) equations with the relevant EOS as input,
which embodies the theoretical information of our theory on dense matter.
The hybrid EOS including both HM and QM is usually obtained
by combining EOSs of HM and QM within individual theories/models.
Unfortunately, while the microscopic theory of the nucleonic EOS has
reached a high degree of sophistication
\cite{Glen00,baldo97,zhou04,Schulze06,Li08zh},
the QM EOS is still poorly known at zero temperature and
at the high baryonic density appropriate for NS.
A lot of work has been done to go beyond the MIT bag model, e.g.,
using perturbative QCD
\cite{Baluni78,Fraga01,Kurkela10,Xu15},
the density-dependent-quark-mass model
\cite{Fowler81,Chakrabarty91,Benvenuto95,Li10,Torres13},
the Nambu-Jona-Lasinio model
\cite{Klahn07,Buballa05,Schertler99,Klahn13,Chu16},
the chiral-quark-meson model \cite{Zacchi15},
or the quasi-particle model
\cite{Peshier00,Tian12,Zhao15a,Li15}.
However,
the EOS remains poorly known due to the nonperturbative character of QCD.

The Dyson-Schwinger equations (DSE) provide a continuum
approach to QCD that can simultaneously address both confinement and
dynamical chiral symmetry breaking \cite{Roberts94dr,Alkofer01}.
They have been applied with success to hadron physics in vacuum
\cite{Maas13,Roberts07jh,Fischer09aa,Chang09,Chang13,Eichmann10,Eichmann16}
and to QCD at nonzero chemical potential and temperature
\cite{Maas13,Roberts00aa,Chen08,Fischer09b,Qin10,Gao14,Fischer14a,Fischer14b,
Klahn15,Eichmann16b,Gao16,Muller13,Muller16,Zhao15}.
Both MIT and NJL model have been recognized
as limiting cases of the DSM \cite{Klahn13,Klahn15}.

In this paper,
we use a DSM for QM based on our previous work \cite{Chen11,Chen15a,Chen16},
where we have investigated the hadron-quark phase transition from HM to 3QM
in compact stars and the structure of hybrid stars with 3QM,
in combination with a nuclear-matter EOS within the Brueckner-Hartree-Fock
(BHF) many-body approach \cite{baldo97,zhou04,Schulze06,Li08zh}.
The possibility of SQM and SQS was also investigated in our model,
with a full scanning of the parameter space \cite{Chen16}.
However, there are still free parameters due to uncertainties
of the gluon propagator and vacuum pressure in our model.
Especially, there is an ambiguity with the parameter $\bds$
when including strange quarks.
To complement the above studies,
we will investigate in this work the phase transition from HM to 2QM
and hybrid stars with 2QM,
also in combination with the HM EOS within the BHF many-body approach.
Particular attention will be paid to the constraint on the maximum mass of NS,
$\mmax > 2\ms$,
which cannot be fulfilled by SQSs,
but by hybrid stars with 3QM or 2QM in our model.

The paper is organized as follows.
In section~2 we briefly discuss the HM EOS within the BHF approach
and the DSM for QM.
In section~3 we analyze the phase transition from HM to 2QM,
and present results on the structure of two-flavor hybrid stars,
with detailed comparison with the 3QM case.
Section~4 contains our summary and conclusions.

\section{Formalism: EOS of dense matter}
\label{s:eos}

\subsection{Hadronic matter within Brueckner theory}

Our EOS of HM obtained within the BHF approach \cite{Baldo99}
has been amply discussed in previous publications \cite{Chen11}.
The basic input quantities of the calculation
are the nucleon-nucleon two-body potentials, namely
Argonne $V_{18}$ \cite{Wiringa95},
Bonn~B \cite{Machleidt87,Machleidt89},
or Nijmegen~93 \cite{Nagels78,Stoks94},
supplemented with compatible three-body forces
\cite{Li08zh,Lejeune86,Zuo02,Li08,Carlson83,Schiavilla86,Pudliner97},
such that the empirical saturation properties of nuclear matter
are well fulfilled.
The different high-density behavior of the EOSs leads nevertheless to
different predictions for NS maximum masses and radii, in particular
\cite{Li08zh,Taranto13}.

The consistency of various BHF EOSs with further experimental constraints
\cite{Miskowiec94,Fuchs06,Danielewicz02}
was also studied in detail in \cite{Li08zh,Taranto13}.
In this work we choose the nucleonic EOS with the Argonne $V_{18}$ potential,
combined with a compatible microscopic three-body force \cite{Li08zh},
which gives a hard EOS
in agreement with empirical constraints up to high density,
and a large NS maximum mass of about 2.3$\ms$.

This approach has also been extended with inclusion of hyperons
\cite{sch98,baldo98,baldo00,vi00,Schulze06},
which might appear in the core of a NS.
The hyperonic EOS in this theory is very soft,
which results in too low maximum masses of NS \cite{Schulze11}.
Furthermore, in combination with our DSM,
either the phase transition to QM
leads to a too soft EOS and low NS maximum mass,
or the QM onset is suppressed by the hyperons \cite{Chen11}.
However, required input information like hyperon-hyperon potentials or
hyperonic three-body forces is currently completely unknown,
such that firm conclusions cannot be drawn.
We therefore prefer to not address this issue in our work that is focused
on the hadron-quark phase transition.

The BHF calculations provide the energy density $\eps$ of the bulk system
as a function of the relevant partial densities $\rho_i$,
from which all other thermodynamical quantities can be obtained,
in particular chemical potentials and pressure,
\be
 \mu_i = {\partial \eps \over \partial \rho_i} \:,
\ee
\be
 p(\rob) = \rob^2 {d\over d\rob} {\eps\over\rob}
 = \rob {d\eps \over d\rob} - \eps
 = \rob \mu_B - \eps \:.
\ee
The parameterized energy density function can be found in Ref.~\cite{Li08zh}.

\subsection{Quark phase with the Dyson-Schwinger model}
\label{s:qm}

For cold dense QM,
we adopt a model based on the DSE of the quark propagator,
described in detail in our previous papers \cite{Chen11,Chen15a,Chen16}.
In the following, we only give a brief introduction to the model.
We start from the gap equation for the quark propagator $S(p;\mu)$
at finite chemical potential $\mu$,
\be
 \Sigma(p;\mu) =
 \int\!\!\! \frac{d^4q}{(2\pi)^4}
 g^2(\mu) D_{\rho\sigma}(p-q;\mu)
 \frac{\lambda^a}{2} \gamma_\rho S(q;\mu)
 \Gamma^a_\sigma(q,p;\mu) \:,
\label{gensigma}
\ee
where $\lambda^a$ are the Gell-Mann matrices,
$g(\mu)$ is the coupling strength,
$D_{\rho\sigma}(k;\mu)$ the dressed gluon propagator,
and $\Gamma^a_\sigma(q,p;\mu)$ the dressed quark-gluon vertex
at finite chemical potential.
To solve the equation,
one requires an ansatz for both $D_{\rho\sigma}$ and $\Gamma^a_\sigma$.
In our model,
the ansatz for $D_{\rho\sigma}$ and $\Gamma_\sigma$ is parameterized as
\be
 g^2 D_{\rho \sigma}(p-q) \Gamma_\sigma^a(q,p) =
 {\cal G}(k) \, D_{\rho\sigma}^{\rm free}(k)
 \frac{\lambda^a}{2}\Gamma_\sigma(q,p) \:,
\label{KernelAnsatz}
\ee
wherein
$D_{\rho\sigma}^\text{free}(k\equiv p-q) =
(\delta_{\rho\sigma}-\frac{k_\rho k_\sigma}{k^2})\frac{1}{k^2}$
is the Landau-gauge free gluon propagator,
$\Gamma_\sigma(q,p)$
represents the tensor structure of the quark-gluon vertex ansatz,
while other dressing effects of the vertex are assumed to depend
only on the gluon momentum $k$ and,
together with the dressing of the gluon propagator,
are included in a model effective interaction ${\cal G}(k)$.

For $\Gamma_\sigma$, we use the rainbow approximation,
i.e., the bare vertex form
$\Gamma_\sigma=\gamma_\sigma$.
For the effective interaction,
we employ an infrared-dominant interaction
modified by the quark chemical potential \cite{Chen11,Jiang13}
\be
 {\cal G}(k) =
 4\pi^2 d \frac{k^4}{\omega^6} e^{-\frac{k^2+\al\mu^2}{\omega^2}} \:.
\label{gaussiangluonmu}
\ee
The parameters $\omega$,$d$ in Eq.~(\ref{gaussiangluonmu})
are discussed in \cite{Alkofer02,Chang09}:
$\omega$ represents the energy scale in nonperturbative QCD,
like $\Lambda_\text{QCD}$,
and $d$ controls the effective coupling strength.
Their values as well as the quark masses are obtained
by fitting light ($\pi$ and $K$) meson properties
and the chiral condensate in vacuum \cite{Alkofer02,Chang09},
and we use the set
$\omega=0.5\;\text{GeV}$ and $d=1\;\text{GeV}^2$.
We choose the quark masses
$m_{u,d}=0$ and $m_s=115\;\text{MeV}$.

The phenomenological parameter $\al$ is of particular importance in our work,
since it represents a reduction rate of the
effective interaction with increasing chemical potential.
However, it cannot yet be fixed independently.
Obviously, $\al=\infty$ corresponds to a noninteracting system
at finite chemical potential,
i.e., a simple version of the MIT bag model;
in the following we call that case the MIT limit.
In previous and present work
we investigate the full parameter space $0<\al<\infty$.

All the relevant thermodynamical quantities of cold QM can be computed
from the quark propagator at finite chemical potential,
except a boundary value of the pressure $P$,
which is represented by a phenomenological bag constant $\bds$,
\be
 P(\mu_u,\mu_d,\mu_s) = - \bds + \sum_{q=u,d,s}
 \int_{\mu_q^0}^{\mu_q}\! d\mu \,n_q(\mu) \:,
\ee
where the density distributions $n_q$ are obtained from the quark propagator
\cite{Chen11,Chen08,Klahn10} 
\be
 n_q(\mu) = 6 \int\frac{d^3 p}{(2\pi)^3} \, f_q(|\bm{p}|;\mu) \:,
\label{nqmu}
\ee
\be
 f_q(|\bm{p}|;\mu) =
 \frac{1}{4\pi} \int_{-\infty}^\infty \! dp_4 \,
 {\rm tr}_{\rm D}\big[-\gamma_4 S_q(p;\mu)\big] \:,
\label{nqmuf1}
\ee
where the trace is over spinor indices only.

As discussed in \cite{Chen11,Chen16},
$\bds\approx90$\mfm
can be obtained from the vacuum pressure in the massless 2QM case
in our model,
but there are ambiguities when including strange quarks.
In this paper we allow a free variation of $\bds$,
but expecting it to be of the same order as 90\mfm.
More constraints on $\bds$ can be obtained from, e.g.,
the stability of normal symmetric nuclear matter against QM
\cite{Chen16,Klahn15}.
In this paper we mainly investigate the constraints
imposed by the observed NSs with $M > 2\ms$.
In the following, $\bds$ is always given in units of
\mfm\ in the text and figures.

\section{Results and discussion}
\label{s:res}
\subsection{EOS of two-flavor quark matter}
\label{s:2qm}

We investigate in the following NS matter, i.e.,
cold, neu\-trino-free, charge-neutral,
and beta-stable matter \cite{Chen11,Chen12},
characterized by two degrees of freedom,
the baryon and charge chemical potentials $\mu_B$ and $\mu_Q$.
The corresponding equations are
\be
 \mu_i = b_i \mu_B + q_i \mu_Q \ ,\quad
 \sum_i q_i \rho_i = 0 \:,
\ee
$b_i$ and $q_i$ denoting baryon number and charge of the particle species,
$i=n,p,e,\mu$ in the nuclear phase
and $i=u,d,(s),e,\mu$ in the quark phase, respectively.

In Fig.~\ref{f:eosmu} we first illustrate the corresponding
EOS $P(\mu_B)$ (lower panel)
and the baryon number density $\rho_B(\mu_B)$ (upper panel)
of 2QM and 3QM in our DSM with $\bds=90$
and various values of the parameter $\al$.
For comparison, we also show the results
for hadronic nuclear matter from the BHF V18 EOS \cite{Li08zh}
(thick solid black curve).
More detailed results of the 3QM EOS can be found in Ref.~\cite{Chen11}.

\vskip-10mm
\centerline{\includegraphics[scale=0.45]{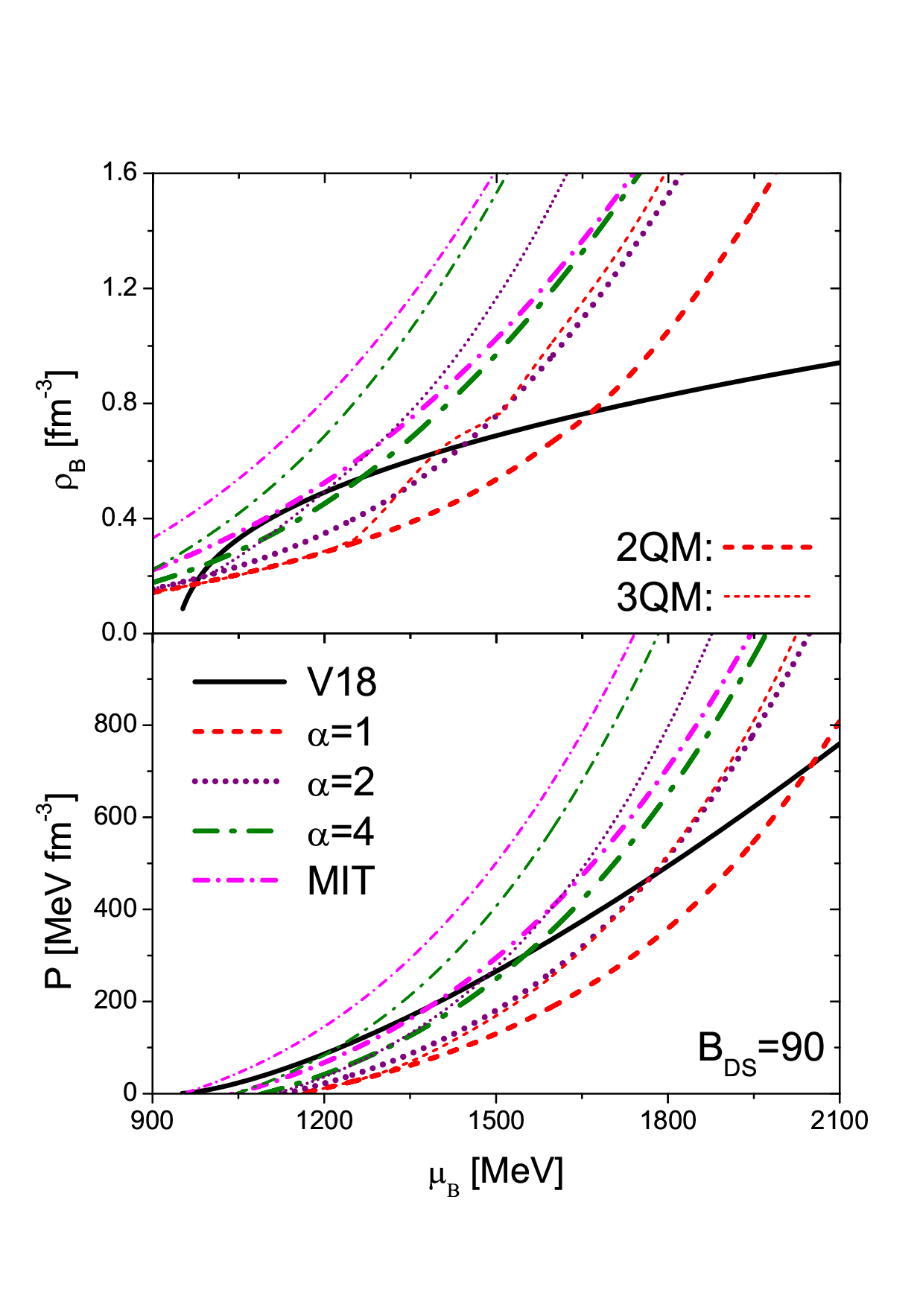}}
\vspace{-10mm}
\figcaption{\label{f:eosmu} 
Pressure (lower panel)
and baryon density (upper panel)
vs.~baryon chemical potential of dense NS matter
for different models and parameters.
Results for 2QM/3QM are shown as thick/thin curves.}

Under a Maxwell construction,
the physically realized phase
at a given chemical potential $\mu_B$
is the one with the highest pressure,
and the crossing points of the nuclear and quark pressure curves
represent the transition between HM and QM phases.
Obviously, at a fixed chemical potential,
the density and pressure are larger for 3QM than for 2QM.
Therefore, the Maxwell phase transition to 2QM is located at
larger chemical potential than that to 3QM.
On the other hand, at a fixed baryon density,
the baryon chemical potential of 3QM is smaller than that of 2QM,
and consequently, the energy density of 3QM is smaller than that of 2QM.
Therefore, the Maxwell phase transition to 2QM is located
at larger density than that in the 3QM case.
For example, one can see that in the $\al=\infty$, i.e., the MIT(2QM) limit,
the Maxwell phase transition is located at $\rho_B=0.62$\fm3 in HM,
which is much larger than that in the MIT(3QM) case, $\rho_B=0.10$\fm3.

With decreasing $\al$
(increasing interaction strength),
the curves of the pressure and density of QM shift downwards,
and correspondingly the phase transition points shift to larger densities.
With an unscreened interaction ($\al=0$) no phase transition at all is possible.
When $\al$ is not large enough,
the phase transition can be at densities higher than 1.01\fm3,
which corresponds to the central density of a NS
with maximum mass of 2.34$\ms$,
and hybrid NSs with Maxwell construction
cannot be built for such parameter choices.

\vspace{-2mm}
\begin{center}
\includegraphics[scale=0.4]{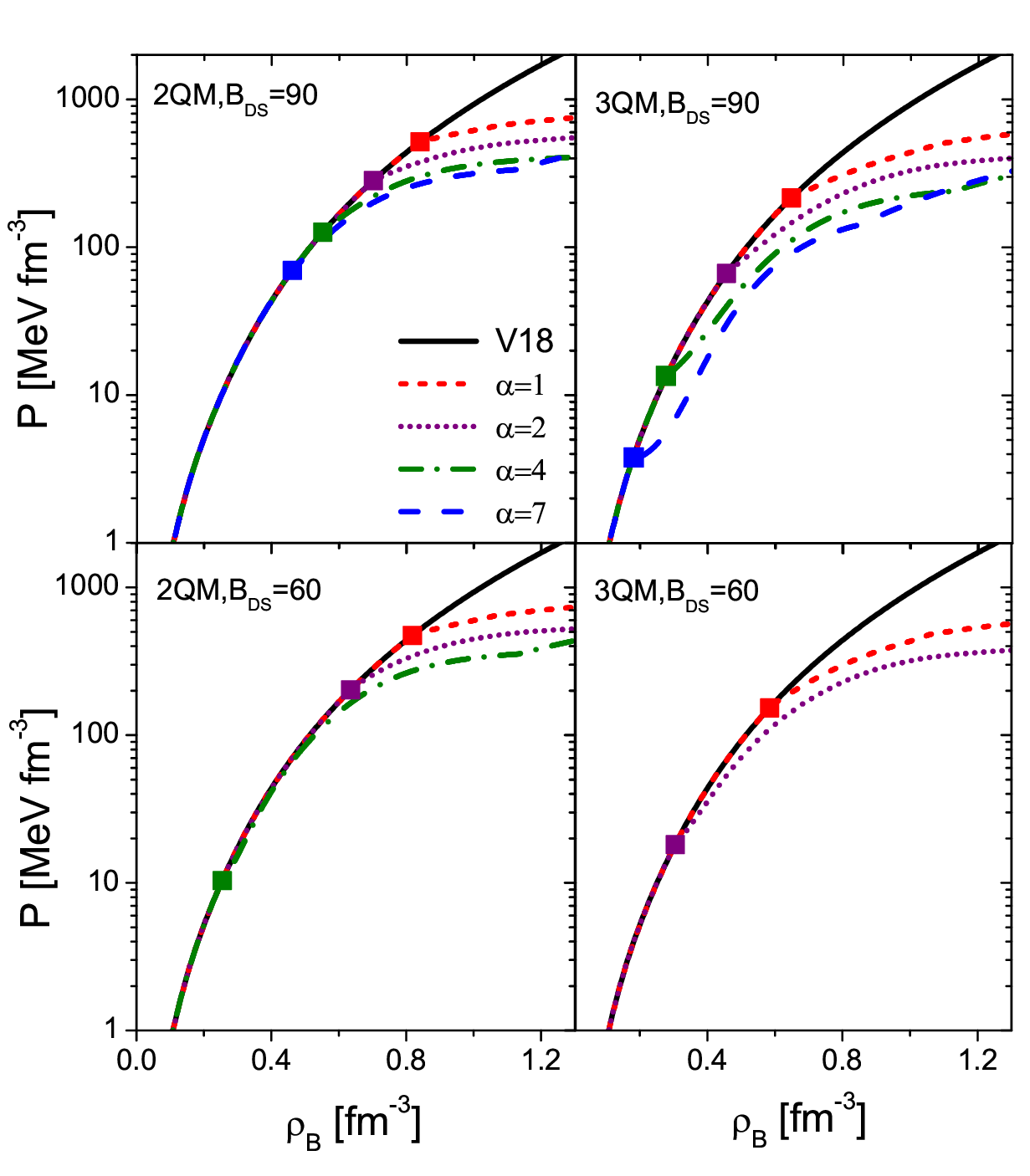}
\vspace{-1mm}
\figcaption{\label{f:prho}  
NS matter pressure vs.~baryon number density
for different EOSs with
2QM (left panels), 3QM (right panels),
$\bds=90$ (upper panels), and $\bds=60$ (lower panels).
Markers indicate the onset of the Gibbs phase transition.
In the lower panels, some curves corresponding to too low maximum masses
of hybrid stars are not shown.}
\end{center}

Considering varying the parameter $\bds$,
it is obvious that with increasing (decreasing) $\bds$,
the curves in the lower panel move downwards (upwards),
while those in the upper panel remain fixed.
Therefore, the phase transition densities increase (decrease) correspondingly.
It is worth noting that when $\al$ is small,
the phase transition occurs at high density,
and is fairly insensitive to the change of $\bds$,
which is not the case for large $\al$.

Regarding the difference between 2QM and 3QM results,
at given parameters $\bds$ and $\al$,
the latter QM is obviously more bound,
i.e., the pressure is lower.
However, readjusting the parameters, similar results can be obtained,
see, e.g., the 2QM $\al=1$ and the 3QM $\al=2$ results
in Fig.~\ref{f:eosmu}.

In Fig.~\ref{f:prho} we show the results of pressure vs.~density,
for phase transitions from HM to 2QM (left panels) or 3QM (right panels)
under the more realistic Gibbs construction \cite{Glen92,Maru07,Glen00,Yas13}
with $\bds=90(60)$ in the upper(lower) panel.
When the phase transition is at high density
(small value of $\al$),
where the pressure is large,
the relative influence of $\bds$ on the EOS is small.
However, when the phase transition is at low density
with a large value of $\al$,
it is quite sensitive to $\bds$.
For example, the onset density 
of the Gibbs phase transition to 2QM
shifts from 0.84\fm3 down to 0.82\fm3 when $\al=1$,
but from 0.55\fm3 down to 0.26\fm3 when $\al=4$.

Qualitatively, the effects of varying the parameters $\al$ and $\bds$
are the same in both 2QM and 3QM cases.
Comparing the difference between 2QM and 3QM with the same
parameters $\al$ and $\bds$,
the phase transition in 3QM is at lower density and pressure,
and the EOS is much softer than in 2QM.
Therefore, the maximum mass of 3QM hybrid stars is smaller than in the 2QM case.
These results are discussed in the following section.

\subsection{Structure of hybrid stars}
\label{s:hs}

As usual, we assume that a NS is a spherically symmetric distribution of
mass in hydrostatic equilibrium and
obtain the stellar radius $R$ and the gravitational mass $M$
by the standard process of solving the TOV equations \cite{Shapiro83}.
We have used as input the EOSs with the Gibbs construction discussed above
and shown in Fig.~\ref{f:prho}.
For the description of the NS crust,
we have joined the hadronic EOS with the
ones by Negele and Vautherin \cite{Negele73} in the medium-density regime,
and the ones by Feynman-Metropolis-Teller \cite{Feynman49} and
Baym-Pethick-Sutherland \cite{Baym71} for the outer crust.

In Fig.~\ref{f:M90} we show the gravitational mass of
2QM (lower panels) and 3QM (upper panels) hybrid stars
vs.~the central baryon density (left panels)
and the radius (right panels)
with various $\al$ and $\bds=90(60)$ [thick(thin) curves].
In comparison we also show the nucleonic NSs (thick solid curve).
The results are in line with those shown in Fig.~\ref{f:prho},
namely increasing the damping parameter $\al$ of the effectively repulsive
QM interaction leads to an earlier onset of the QM phase in NS
and a consequential reduction of the maximum mass.

Obviously, in a certain range of the parameters $\bds$ and $\al$,
there exist 2QM-hybrid stars with $\mmax \geq 2\ms$.
For example, under such a constraint,
$\al$ should be smaller than 7.2(4.3) for $\bds=90(60)$.
The corresponding parameter curve for $\mmax=2\ms$
is shown as the solid black curve in Fig.~\ref{f:Bal}.
We also show the lower boundary of $\bds$ at given $\al$
under the stability constraint of symmetric nuclear matter
against 2QM at the saturation point (dash-dotted green curve),
as well as the upper boundary of $\bds$ for the SQM hypothesis
(dotted red curve),
see details in \cite{Chen16}.
We can see that,
although 2QM hybrid stars can coexist with SQSs
in the same parameter space,
which is enclosed by the dotted red curve and the dash-dotted green curve,
their mass is below 2$\ms$ in that domain.
Therefore,
if SQSs exist and NSs convert first to 2QM-hybrid stars and then to SQSs,
the maximum mass of the metastable 2QM-hybrid stars
(and the SQSs)
is lower than two solar masses in the DS model.

\vspace{-2mm}
\centerline{\includegraphics[scale=0.5]{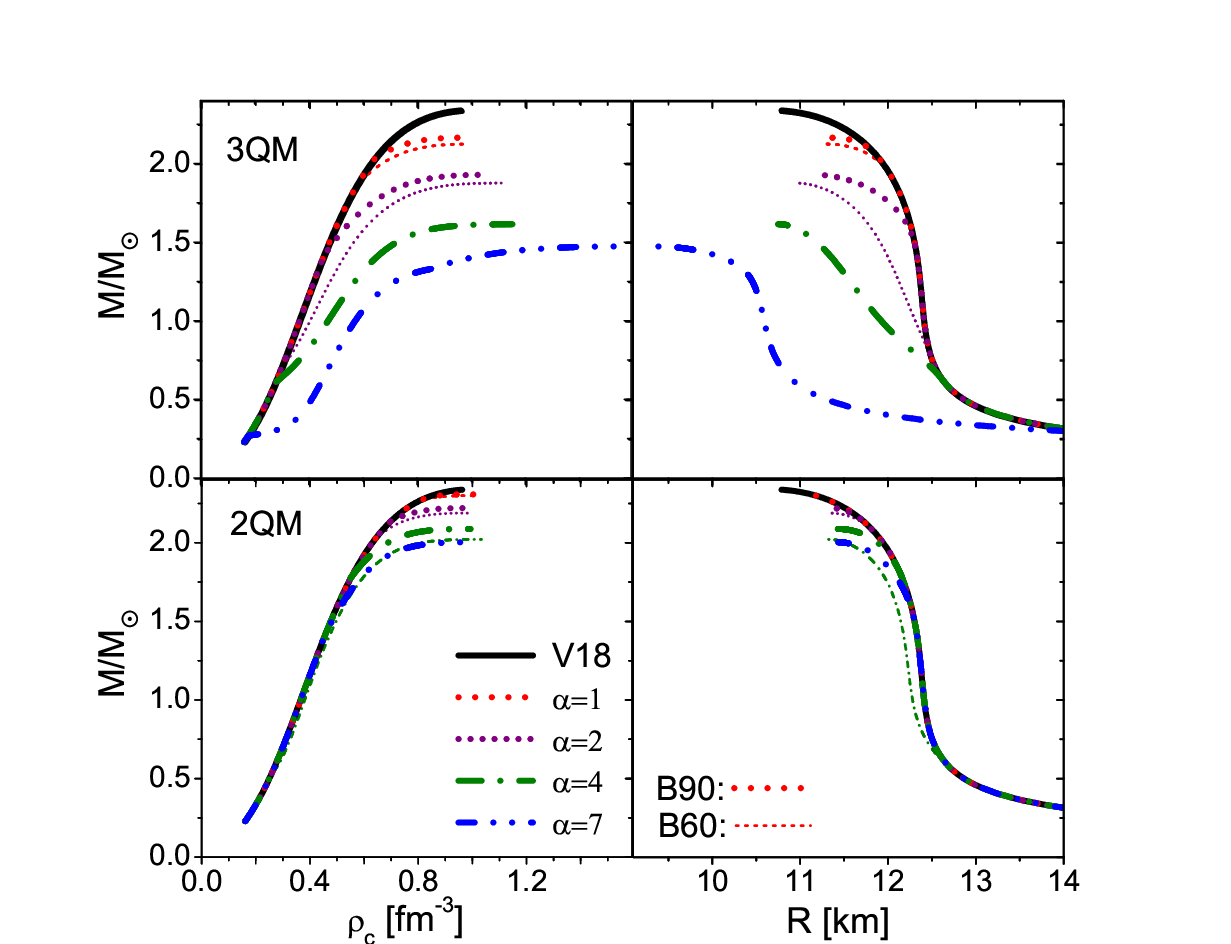}}
\vspace{-3mm}
\figcaption{\label{f:M90}  
NS gravitational mass vs.~central baryon density (left panels)
and radius (right panels),
using EOSs shown in Fig.~\ref{f:prho},
with $\bds=90$ (thick curves) and $\bds=60$ (thin curves).}

It is also worth to compare with 3QM-hybrid stars \cite{Chen11}.
The maximum mass of 2QM-hybrid stars can reach a higher value
with the same parameters in our model,
because the phase transition occurs at higher density
and correspondingly in heavier stars.
In other words, under the constraint of $\mmax=2\ms$,
the corresponding (dashed blue) parameter curve for 3QM lies at the top/right
of that for 2QM (solid black) in Fig.~\ref{f:Bal}.
For 3QM-hybrid stars with the constraint $\mmax \geq 2\ms$,
$\al$ should be smaller than 1.65(1.45) for $\bds=90(60)$,
see the markers in  Fig.~\ref{f:Bal}.

\vspace{-5mm}
\begin{center}
\includegraphics[scale=0.3]{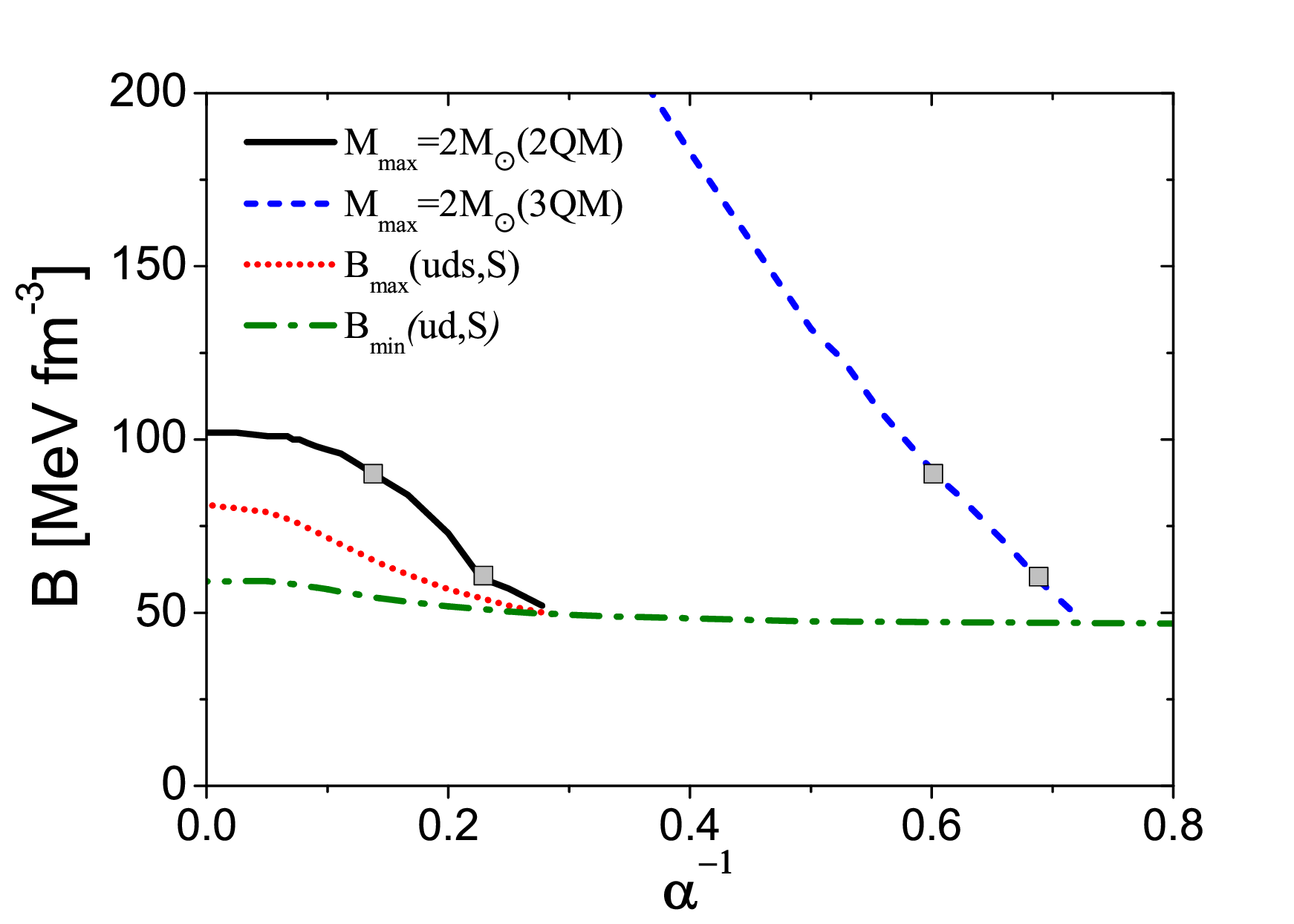}
\vspace{-2mm}
\figcaption{\label{f:Bal}  
The parameters space in the ($\bds,\al^{-1}$) plane
that allows hybrid stars with $\mmax\geq 2\ms$,
which lie on the top/right side of the corresponding
solid black or dashed blue curves.
The dash-dotted green curve represents the lower boundary of $\bds$
under the stability constraint of symmetric nuclear matter
against 2QM at the saturation point,
and the dotted red curve represents the upper boundary of $\bds$
for the SQM hypothesis, see details in \cite{Chen16}.
The markers indicate parameter sets chosen in Fig.~\ref{f:MQM}.}
\end{center}

\subsection{Quark matter content and energy release}

It is interesting to investigate how much QM can be present in hybrid stars.
As an illustration we show in Fig.~\ref{f:MQM}
the QM fractions $M_\text{QM}/M$ in hybrid stars,
for several NS configurations with $\mmax=2\ms$
(see markers in Fig.~\ref{f:Bal}).
It can be seen that with decreasing $\bds$,
QM can appear in lighter NSs and the mass fraction can be much higher,
up to nearly 30\% of 2QM with $\bds=60$,
which is close to the minimal value according to Fig.~\ref{f:Bal}.
When $\bds$ is large, e.g., $\bds=90$,
the curves corresponding to 2QM are close to the 3QM case,
i.e., the masses of QM in hybrid stars are insensitive to the components of QM.
However, when $\bds=60$,
the 2QM can appear in lighter NSs and the masses reach larger values
than in the 3QM case.
For example, even in a 1.4 solar mass NS,
the mass fraction of 2QM can be as large as 18\%.
The results shown in Fig.~\ref{f:MQM} represent upper limits on
the quark fraction in hybrid stars within the DS+V18 model.
Choosing parameter sets with $\mmax>2\ms$,
the onset of the QM phase will be delayed and the QM fraction reduced.

The hybrid stars in the DS model are lighter than nucleonic NSs
with the same baryonic mass.
Therefore it is possible that a NS converts to a hybrid star
with a phase transition,
accompanied by an energy release.
As an illustration we show in the lower panel of Fig.~\ref{f:MQM}
the mass difference of NSs and hybrid stars with the same baryonic mass.
One expects that that quantity depends strongly
on the QM fraction in hybrid stars,
and in fact the upper and lower panels of Fig.~\ref{f:MQM}
show qualitatively similar results.
Quantitatively, the maximum value of the mass difference
can reach about 7 promille of the solar mass,
with 2QM in the core and the smallest possible $\bds$.
The corresponding energy release is about $10^{52}\;$erg
in the conversion from NSs to hybrid stars,
which is much smaller than the possible energy release in the conversion
from NSs to SQSs \cite{Chen16}.
With increasing of $\bds$ and/or the maximum mass value of hybrid stars,
the mass difference and corresponding energy release decrease
along with the fraction of QM in hybrid stars.

\vspace{-21mm}
\begin{center}
\includegraphics[scale=0.3]{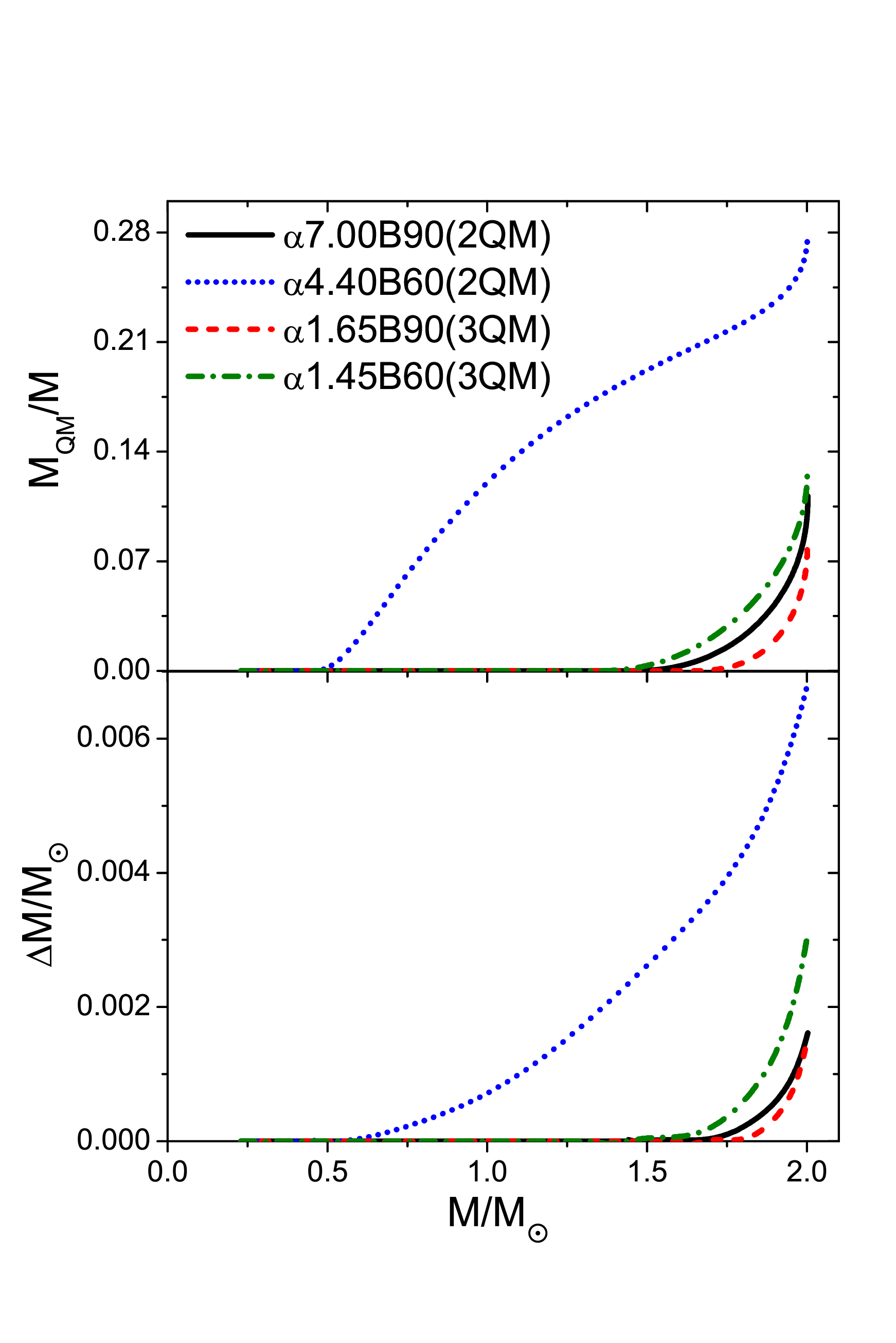}
\vspace{-12mm}
\figcaption{\label{f:MQM}  
Upper panel:
The mass fractions of 2QM or 3QM in hybrid stars with $\mmax= 2\ms$
for different parameter sets of $\bds$ and $\al$,
as indicated in Fig.~\ref{f:Bal}.
Lower panel:
The corresponding mass difference of NSs and hybrid stars
with the same baryonic masses.}
\end{center}

\section{Conclusions}
\label{s:sum}

We have investigated the phase transition from HM to 2QM
and hybrid stars with 2QM in our DSM for QM,
in combination with a nuclear matter EOS within the BHF many-body approach.

In the full space of parameter $\al$ and $\bds$,
we investigate the EoS of 2QM and the phase transition from HM to 2QM.
In comparison with 3QM with the same parameters,
the phase transition from HM to 2QM locates at higher chemical potential and density.
Correspondingly, the hybrid EoS with HM and 2QM mixed phase are stiffer than that in the 3QM case,
 and the 2QM hybrid stars have larger maximum mass.
On the other hand, we show the allowed parameter space in the 2QM case under the constraint of $\mmax>2\ms$,
 which is much larger than that in the 3QM case.

This study completes previous works on 3QM hybrid stars.
2QM is less bound than 3QM and could only exist as a
transitory phase during the stellar evolution.
After that, either a 3QM hybrid NS is formed or collapse to a black hole occurs.

We have shown that for parameter choices that respect the
$\mmax>2\ms$ condition,
the possible quark matter content of 2QM(3QM)-hybrid NSs
is limited to about 30\%(15\%),
due to constraints on the onset density of QM in the stellar matter.
Such heavy 2QM hybrid stars would eventually collapse to black holes,
because the compatible 3QM stars with the same parameters $\al$ and $\bds$
are too strongly bound with a much smaller maximum gravitational mass and baryon mass.
The related energy release during the HM-QM transition is limited
to about $10^{52}\;$erg.
We have only considered energy balances
and disregarded the dynamical development of that transition,
which is still a difficult theoretical problem.

These results obviously depend on the chosen nucleonic EOS
and the parameters in the DSM model.
For the future it will be important to establish direct estimates of
the DS model parameters in a more fundamental way from QCD,
and to provide in this way more reliable predictions,
and also to clarify the qualitative differences between
different current theoretical quark models.

\section*{Acknowledgments}

We acknowledge
financial support from NSFC (11305144, 11475149, 11303023),
Central Universities (CUGL 140609) in China.
Partial support comes from ``NewCompStar," COST Action MP1304.

\vspace{5mm}

\end{multicols}
\end{document}